\documentstyle[aps,preprint,tighten,prb]{revtex}
\begin{document}
\preprint{submitted to Phys. Rev. B}
\draft
\title{Statistical properties of level widths and
conductance peaks in a quantum dot}
\author{E. R. Mucciolo$^{\dagger}$, V. N. Prigodin$^{\ddagger}$, and
B. L. Altshuler$^{\dagger}$}
\address{$^{\dagger}$Department of Physics, Massachusetts Institute of
Technology, Cambridge, Massachusetts 02139 \\
and \\
$^{\ddagger}$Max-Planck-Institut f\"ur Festk\"orperforschung,
Postfach 80 06 65, D-70506 Stuttgart, Germany}
\date{June 10, 1994}
\maketitle
\begin{abstract}
We study the statistics of level widths of a quantum dot with extended
contacts in the absence of time-reversal symmetry. The widths are
determined by the amplitude of the wavefunction averaged over the
contact area. The distribution function of level widths for a
two-point contact is evaluated exactly. The distribution resembles
closely the result obtained when the wavefunction fluctuates
independently at each point, but differs from the one-point case.
Analytical calculations and numerical simulations show that the
distribution for many-point contacts has a power-law behavior at small
level widths. The exponent is given by the number of points in the
lead and diverges in the continuous limit. The distribution of level
widths is used to determine the distribution of conductance peaks in
the resonance regime. At intermediate temperatures, we find that the
distribution tends to normal and fluctuations in the height of the
peaks are suppressed as the lead size is increased.
\end{abstract}

\section{Introduction}
\label{sec:introduction}

Usually any measurement of conductance in a metallic system assumes at
least two attached leads. The existence of these leads connecting the
system to reservoirs broadens the electronic energy levels. For a
macroscopic sample, the width of an energy level is typically much
larger than the distance between neighboring levels. As a result, one
observes a smooth dependence of the conductance on the Fermi energy.

Recently, the advances in nanometer technology have made possible the
fabrication of very small semiconductor devices, known generically as
quantum dots, \cite{kirk92,kastner92,grabert92} where one has a fixed
number of conduction electrons confined into an island. In these
systems, one can narrow the level width by reducing either the size of
the leads or their transmittance. In the first case it is customary to
speak about channels: If we consider an ideal lead as a wave guide
with a given cross section, we can associate a channel to each state
due to transverse quantization. The number of channels is
approximately the area, in units of electron wavelength, of the
contact between lead and dot and each channel is characterized by its
transmittance.

The conductance of a system can be calculated through the well-known
Landauer-B\"uttiker formula. \cite{landauer,buettiker} Following this
approach, the distribution of conductances of a quantum dot has been
evaluated for two distinct cases: for weakly coupled, pointlike, leads
\cite{prigodin93} of a closed system, and for leads with any number of
channels attached to a ballistic cavity.
\cite{baranger94,jalabert94}

Here we will consider the situation at low transmittance, when leads
behave as tunnel contacts. In this case, each electron eigenstate
corresponds to a peak in the Fermi energy (gate voltage) dependence of
the conductance. \cite{kastner92,grabert92} This allows one to make a
real spectroscopy of electrons in quantum dots. The height of each
peak is determined by the probabilities of tunneling through the
leads, as well as by the amplitude of the wavefunction near the
contacts. The later means that these heights are randomly distributed.
The distribution function of the peaks has been determined for the
case of pointlike contacts in Refs.
\onlinecite{prigodin93,jalabert92}, and \onlinecite{stone93}. In this
paper we will deduce the distribution considering leads of arbitrary
size.

We begin by studying the statistics of level widths, which is
connected to the fluctuations of conductance peaks through the
Landauer-B\"uttiker formula. The supersymmetry technique and the
nonlinear $\sigma$ model \cite{efetov83,verbaarschot85} provide the
framework for the exact calculation of the distribution of level
widths for leads with two-point contact. The distribution differs
drastically from the single-point case, but we find that inclusion of
correlation between the wavefunction fluctuations at the two points in
the lead does not significantly alter the overall form of the
distribution of level widths. This analytical result is confirmed
through numerical simulations of a quantum dot. The exact distribution
of conductance peaks is also evaluated for two-point leads and we find
that it deviates from the single-point distribution mainly for small
peak heights, where it shows a linear dependence.

For leads with many-point contacts we do not know of any method which
enables a general evaluation of the distribution function of level
widths. Thus we assume that the wavefunction at each point fluctuates
independently and proceed to calculate the total distribution. By
comparing numerical with analytical results, we show that this
approximation yields a good qualitative understanding of the
large-lead limit, since correlations do not effect the behavior of the
distribution very strongly.

An important use of quantum dots is found in the study of chaos.
\cite{baranger93,marcus93,weiss93,berry94,keller94,fromhold94}
Because one can obtain extremely clean samples and shape them into
different forms, it is now possible to fabricate the so-called quantum
billiards, where one has a unique chance to study quantum and
semiclassical physics. \cite{baranger93,tomsovic92} Although the
supersymmetry method assumes averaging over disorder, one can
conjecture on very firm grounds (the ergodic hypothesis) that it
applies quite generally to quantum chaos problems. Therefore, we
expect that our results should be able to describe any system where
the underlying dynamics is chaotic, regardless as to whether it is
diffusive or ballistic

The organization of this paper goes as follows. In Sec.
\ref{sec:ideas} we describe the model used and the basic concepts
related to conductance peaks and level widths. The way the numerical
simulations were done is explained in Sec. \ref{sec:numerics}. Some of
the results found in Ref. \onlinecite{prigodin93}, which initially
motivated the present work, are reviewed in Sec. \ref{sec:one-site}.
The case of two-point leads is treated exactly in Sec.
\ref{sec:two-site} and the case of many-point leads, under the
approximation of independent point fluctuations, is left to Sec.
\ref{sec:n-site}. Finally, in Sec. \ref{sec:conclusion} we draw our
conclusions and point out some experimental consequences of our work.

\section{Formulation of the problem}
\label{sec:ideas}

The electronic system we consider is formed by noninteracting spinless
fermions confined to a small region through some strong confining
potential. The system is probed by two leads, described by discrete
sets of points, which are weakly coupled to the system. More
precisely, we have the Hamiltonian
\cite{prigodin93,zirnbauer92}
\begin{equation}
H = -\frac{\nabla^2}{2m} + U(r) +
\frac{i}{2\pi{\cal N}}\left[\alpha_1 \sum_{r\in A_1} \ \delta(r-r_1)
+ \alpha_2 \sum_{r\in A_2} \ \delta(r-r_2)\right] \ ,
\label{eq:e1.1}
\end{equation}
where $U(r)$ is the potential (containing the effect of random
impurities and confining walls) and ${\cal N}=1/(V\Delta)$ is the mean
density of states, with $V$ denoting the total volume and $\Delta$ the
average level spacing. The term inside the brackets represents the
contact leads, denoted by $A_{1,2}$, and $\alpha_{1,2}$ are the
dimensionless coupling parameters.

In a situation characteristic of scattering experiments, when
$\alpha_{1,2}\gg1$ and the leads are open, the number of channels is
equal to the number of wavelengths which fit into the lead cross
section. As we shall demonstrate, for the weakly coupled leads of a
{\it closed} quantum dot, the distinction between pointlike (single
channel) and multichannel leads is more subtle and deserves a careful
analysis.

When tunneling through the leads in totally suppressed
($\alpha_{1,2}=0$) the energy levels have zero intrinsic width. The
levels will gain a small but finite width if the coupling to the leads
is made very weak, i.e., if $0<\alpha_{1,2}\ll 1$. As one can show by
first order perturbation theory, the broadening will be caused by
local fluctuations of the electron wavefunction in the regions of
contact with the leads. Calling the total level width $\Gamma_\nu$, we
have
\begin{equation}
\Gamma_\nu = \gamma_{\nu,1}+\gamma_{\nu,2} \ ,
\label{eq:e1.2}
\end{equation}
with the partial level widths given by
\begin{equation}
\gamma_{\nu,i} = \alpha_i\left(\frac{V\Delta}{\pi}\right)
\sum_{r\in A_i} |\psi_\nu (r)|^2 \ \ \ \ \ \ (i=1,2) \ ,
\label{eq:e1.3}
\end{equation}
where $\psi_\nu$ is a single-particle eigenfunction of the Hamiltonian
in Eq. (\ref{eq:e1.1}) for $\alpha_{1,2}=0$.

The knowledge of the statistical properties of $\gamma_i$ is crucial
to describe some important and measurable effects in quantum dots.
{}From the distribution of level widths we can extract the distribution
of conductance peaks in the resonance regime. It is rather simple to
understand why: Let us assume that the conductance at a given energy
$E$ can be evaluated through the Landauer-B\"uttiker formula,
\cite{landauer,buettiker}
\begin{equation}
G(E) = \frac{2e^2}{h} \frac{\alpha_1 \alpha_2}{(\pi {\cal N})^2}
\sum_{r_1 \in A_1} \sum_{r_2 \in A_2} \left| \sum_\nu
\frac{\psi_\nu^\ast (r_1) \psi_\nu (r_2)}{E-\varepsilon_\nu +
i\Gamma_\nu/2} \right|^2 \ ,
\label{eq:e1.4}
\end{equation}
where $\varepsilon_\nu$ is the eigenvalue associated with the
eigenfunction $\psi_\nu$. In the resonance regime
($\gamma,T\ll\Delta$), only the eigenstate whose energy is the closest
to $E$ significantly contributes to the conductance. $G(E)$ is very
small when $E$ is between adjacent energy levels, and grows rapidly
when $E\rightarrow\varepsilon_\nu$. Therefore, at a peak of the
conductance, we have
\begin{equation}
G_\nu = \frac{2e^2}{h} \frac{4\alpha_1 \alpha_2}{(\pi
{\cal N}\Gamma_\nu)^2} \sum_{r_1 \in A_1} \sum_{r_2 \in A_2}
|\psi_\nu (r_1)|^2 |\psi_\nu (r_2)|^2 \ .
\label{eq:e1.5}
\end{equation}
Using the last equation, plus Eqs. (\ref{eq:e1.2}) and
(\ref{eq:e1.3}), we obtain the Breit-Wigner formula
\begin{equation}
G_\nu = \frac{2e^2}{h} \frac{4 \gamma_{\nu,1} \gamma_{\nu,2}}
{(\gamma_{\nu,1} + \gamma_{\nu,2})^2} \ ,
\label{eq:e1.6}
\end{equation}
which is correct at zero temperature, or when
$T\ll\alpha_{1,2}\Delta$. In order to describe a wider range of
temperatures, we have to use instead the well-known relation
\begin{equation}
G_\nu = \frac{2e^2}{h} \int \frac{d\epsilon}{4T} \
\frac{\gamma_{\nu,1} \gamma_{\nu,2}} {\epsilon^2 +
(\gamma_{\nu,1} + \gamma_{\nu,2})^2/4} \cosh^{-2}
\left(\frac{\epsilon}{2T}\right) \ .
\label{eq:e1.7}
\end{equation}

Let $N$ be the number of points in each lead. When the leads are
placed very far apart, the connection between the distribution of
level widths, $P_N(\gamma_{\nu,i})$, and the distribution of
conductance peaks, $R_N(g_m)$, is given by the convolution
\begin{equation}
R_N(g_m) = \int_0^\infty d\gamma_{\nu,1} P_N(\gamma_{\nu,1})
\int_0^\infty d\gamma_{\nu,2} P_N(\gamma_{\nu,2}) \ \delta \left(
g_m - G_\nu h/2e^2 \right) \ .
\label{eq:e1.9}
\end{equation}

{}From Eq. (\ref{eq:e1.7}) we can extract another limit, which occurs
when the temperature exceeds the intrinsic energy level width,
$T\gg\alpha_{1,2}\Delta$, but is still low enough so that there are
well-resolved resonance peaks ($T\ll\Delta$). In fact, this
intermediate regime is typical of some experiments \cite{marcus93} and
is described by the Hauser-Feshbach formula
\begin{equation} G_\nu =
\frac{2e^2}{h} \left(\frac{\pi}{2T}\right)
\frac{\gamma_{\nu,1}\gamma_{\nu,2}}{\gamma_{\nu,1} + \gamma_{\nu,2}} \ .
\label{eq:e1.8}
\end{equation}

In practice, one can generate many peaks in the conductance by varying
the energy $E$ (i.e., by sweeping the gate voltage in the quantum
dot), or, else, by varying some external parameter, say, an applied
magnetic field. Similarly, in numerical calculations one can apply a
Aharonov-Bohm flux $\phi$ to the system, and then generate a family of
curves $\varepsilon_\nu(\phi)$, which yields a random sequence of
conductance peaks, $G_\nu(\phi)$. As long as the system is
ergodic, all these procedures are equivalent to the construction of an
ensemble of different realizations of disorder (in the case of a
diffusive regime) or boundaries (if the regime is ballistic).

\section{Numerical simulation}
\label{sec:numerics}

Before we move to the analytical calculations, let us describe the
numerical simulation of a disordered quantum dot which we used to
illustrate the main points of our work. The numerical results will be
shown along the discussion of the analytical ones. Our motivation was
to obtain the histogram $P_N(\gamma)$ as a function of the separation
between the points within the lead. We used a two-dimensional Anderson
model of noninteracting spinless electrons with nearest-neighbor
hopping and diagonal disorder, i.e., the Hamiltonian
\begin{equation}
H_A = -\sum_{\langle i,j \rangle} ( c_i^{\dagger} c_j + c_j^{\dagger}
c_i ) + \sum_{i=1}^N w_i \ c_i^{\dagger} c_i \ ,
\label{eq:e2.1}
\end{equation}
with $w_i$ uniformly distributed in the interval $[-W/2\ ,W/2]$. We
restricted our calculations to only one realization of $\{w_i\}$;
hence, no average over disorder was performed. On the other hand, we
imposed quasiperiodic boundary conditions to the electrons by making
the geometry toroidal and introducing Aharonov-Bohm phases in the
hops,
\begin{equation}
c_i^{\dagger}c_j \longrightarrow c_i^{\dagger}c_j e^{-i\phi_{ij}}
\ ,
\label{eq:e2.2}
\end{equation}
where $\phi_{ij}$ assumes one of two values, $\phi_x$ or $\phi_y$,
depending on whether the hop occurs along the $x$ or $y$ axis,
respectively. By varying $\phi_x$ and $\phi_y$ within the interval
$(0,\pi)$ we obtained a large set of states, thus creating the
statistical ensemble needed to construct the histograms.

For a given set of phases $\phi_x$ and $\phi_y$ the Hamiltonian was
diagonalized through standard procedures and all eigenstates obtained.
The leads were then chosen as sets of sites placed throughout the
grid, such that we could probe the density fluctuations at different
regions of the dot. Notice that the leads were passive objects, which
did not affect the eigenstates.

Briefly, we mention the existence of a technical complication,
intrinsic to simulations of quasiperiodic systems, which is the
crossover with respect to time-reversal breaking. The threshold for
breaking time-reversal symmetry with Aharonov-Bohm fluxes is
proportional to the conductance of the system. \cite{nobuhiko94} If
the disorder is too strong or the system too small, a complete
time-reversal symmetry breaking may never be achieved due to the
periodic nature of fluxes. As a result, in order to fall into the
class of the unitary ensemble and be able to compare the simulations
with the calculations to follow, we needed to work with a large grid
(32$\times$23) and keep the disorder very low. The system was then
maintained marginally diffusive ($W\approx$ 1).

\section{Pointlike leads}
\label{sec:one-site}

We will consider first the simple case of pointlike leads, namely,
when the typical size of the contact is smaller than the area
$\lambda^{d-1}$, where $\lambda$ is the electron wavelength. In the
lattice version of the problem, this certainly happens when the lead
consists of a single site. For a chaotic system with broken
time-reversal symmetry, the amplitude $v = V|\psi_\nu(r)|^2$
fluctuates according to an exponential distribution. This result
follows from the assumption that the components of the eigenfunction
are not correlated among themselves and are uniformly distributed.
\cite{berry91} Therefore, by connecting $v$ to the level width
$\gamma_i$ [see Eq. (\ref{eq:e1.3})] we can write
\begin{equation}
P_1(\gamma_i) = \left(\frac{\pi}{\alpha_i\Delta} \right) \exp
\left(-\frac{\pi\gamma_i}{\alpha_i\Delta} \right) \ .
\label{eq:e3.1}
\end{equation}
(To simplify the notation, we will hereafter drop the eigenstate
label.) In Fig. 1 we have plotted $P_1(\gamma_i)$ obtained from our
numerical simulation against Eq. (\ref{eq:e3.1}) for two different
regions of the spectrum. The states around the bottom of the band used
in the evaluation of the level width distribution were checked to be
extended over the entire grid.

If the distance between the leads $A_1$ and $A_2$ is much larger than
the electron wavelength $\lambda$ (so that the wavefunction
fluctuations at $r_1$ and $r_2$ are independent), we can use Eq.
(\ref{eq:e1.6}) or Eq. (\ref{eq:e1.8}), together with Eqs.
(\ref{eq:e1.9}) and (\ref{eq:e3.1}), to evaluate analytically the
distribution of conductance peaks for single-point leads, $R_1(g_m)$.
This calculation has been recently done in the literature.
\cite{prigodin93,jalabert92} In Ref. \onlinecite{jalabert92} the
authors found an expression for the conductance distribution at
intermediate temperatures not only for single-channel leads, but for
multichannel ones as well. Their derivation, based on random matrix
theory, enabled them to consider the case of time-reversal symmetry,
although they were constrained to treat only the case of independent
channels. In addition, for simplicity, they treated only the situation
of symmetric leads. A similar work is found in Ref.
\onlinecite{stone93}, where the authors have also studied the effect
of chaotic and regular dynamics in the distribution of conductance
peaks.

On the other hand, the authors in Ref. \onlinecite{prigodin93}
performed their analysis exclusively for the case of broken
time-reversal symmetry and pointlike leads, but took into account a
possible asymmetry between the leads. Since we shall extend their
work, we will display and comment on some of their results. For
$T\ll\alpha_{1,2}\Delta$ they obtained the following distribution
\begin{equation}
R_1(g_m) = \frac{\theta(1-g_m)}{2\sqrt{1-g_m}}\frac{[1+(2-g_m)(a^2-1)]}
{[1+g_m(a^2-1)]^2} \ ,
\label{eq:e3.2}
\end{equation}
where the asymmetry parameter $a$ is given by
\begin{equation}
a=\frac{1}{2} \left( \sqrt{\frac{\alpha_1}{\alpha_2}} +
\sqrt{\frac{\alpha_2}{\alpha_1}} \ \right) \ .
\label{eq:e3.3}
\end{equation}
Notice that, according to Eq. (\ref{eq:e3.3}), for $a=1$ (symmetric
leads), the most probable value of $g_m$ is 1. However, for $a>1$ the
distribution begins shifting to lower values of $g_m$. For $a\gg 1$,
the most probable value of $g_m$ tends to zero. This is exactly what
one would expect to happen in a resonance tunneling experiment at zero
temperature: The conductance is maximum when the barriers are
identical.

For the intermediate regime, when $\alpha_{1,2}\Delta\ll T\ll\Delta$,
it is convenient to rescale the distribution to obtain
\cite{prigodin93}
\begin{equation}
{\cal R}_1(x) \equiv \frac{\sqrt{\alpha_1\alpha_2}\Delta }{4T}
R_1(g_m) = x e^{-ax} [ K_0(x) + a K_1(x) ] \ ,
\label{eq:e3.4}
\end{equation}
where
\begin{equation}
x\equiv\frac{4Tg_m}{\sqrt{\alpha_1\alpha_2}\Delta} \ ,
\label{eq:e3.5}
\end{equation}
and $K_n(x)$ is the modified Bessel function of order $n$. Notice that
the rescaling of $g_m$ to $x$ and of $R_1$ to ${\cal R}_1$ turned the
distribution into a temperature-independent function, whose form
depends only on the asymmetry between the leads. We also point out
that for both temperature regimes, one has an finite offset in the
distribution, i.e., $R_1(g_m\rightarrow 0)\ne0$.

\section{Two-point leads}
\label{sec:two-site}

In the previous section we reviewed known results. In this and the
following sections the main goal will be to study the effect of
extended leads in the distribution of level widths and conductance
peaks. We start by analyzing the situation of two-point leads, where
the level width is given by
\begin{equation}
\gamma_i = \frac{\alpha_i\Delta}{\pi}( v_1 + v_2 ) \ .
\label{eq:e4.1}
\end{equation}
The amplitudes are defined as $v_i=V|\psi_\nu(r_i)|^2$, $i=1,2$. When
the points 1 and 2 are far apart ($r=|r_1-r_2|\gg\lambda$), the
amplitudes fluctuate independently and the distribution of level
widths is equal to the convolution of the distributions for the
isolated amplitudes [see Eq.  (\ref{eq:e3.1})]. Consequently,
\begin{equation}
P_2(\gamma_i) = \left(\frac{\pi}{\alpha_i\Delta} \right)^2 \gamma_i \
\exp\left(-\frac{\pi\gamma_i}{\alpha_i\Delta} \right)  \ .
\label{eq:e4.2}
\end{equation}
Notice that the most probable value of $\gamma_i$ is not zero, as for
the one-point case, but $\alpha_i\Delta/\pi$. This fact by itself
signals a strong qualitative change in the distribution. What happens,
then, when we move the points closer together, so that
$r\approx\lambda$? This question can only be fully answered if we know
the joint probability distribution
\begin{equation}
Q_2(v_1,v_2;r) = \left\langle \delta \Bigl( v_1 -
V|\psi_\nu(r_1)|^2 \Bigr) \ \delta \Bigl( v_2 -
V|\psi_\nu(r_2)|^2 \Bigr)
\right\rangle \ .
\label{eq:e4.3}
\end{equation}
However, a simple analysis can be made before one starts seeking the
exact form of $Q_2(v_1,v_2;r)$: Since there is no special reason for
$Q_2(0,0;r)$ to be singular if $r\ne0$, we should have
$P_2(\gamma_i\rightarrow 0)\propto\gamma_i$. On the other hand,
$Q_2(0,0;0)$ is apparently singular in order to Eq. (\ref{eq:e3.1})
hold true.

The exact calculation of $Q_2(v_1,v_2;r)$ can be performed by the
supersymmetry method. \cite{preprint} The result depends on the
symmetry of the Hamiltonian; here we will consider only the case of
broken time-reversal invariance (unitary ensemble), which yields
\begin{equation}
Q_2(v_1,v_2;r) = \frac{1}{1-f(r)^2} \ \exp \left( -\frac{v_1 +
v_2}{1-f(r)^2} \right) \ I_0 \left( \frac{2f(r)}{1-f(r)^2}
\sqrt{v_1v_2} \right) \ ,
\label{eq:e4.4}
\end{equation}
where $f(r)$ is the Friedel-like function
\begin{eqnarray}
f(r) & = & \frac{1}{\nu} \int \frac{dp}{(2\pi)^d} \ e^{-ip\cdot r}
\ \delta(p^2/2m - E) \nonumber \\
&  & \nonumber \\
& = & \cases{ J_0(2\pi r/\lambda) \ \ & $(d=2)$ \cr \cr
(\lambda/2\pi r) \sin(2\pi r/\lambda) \ \ & $(d=3)$ \ , \cr}
\label{eq:e4.5}
\end{eqnarray}
and $I_0(x)$ and $J_0(x)$ are Bessel functions of order zero. The
exact expression for the two-point distribution of level widths can be
readily calculated:
\begin{eqnarray}
P_2(\gamma_i) & = & \int_0^\infty dv_1 \int_0^\infty dv_2 \
Q(v_1,v_2;r) \ \delta \Bigl( \gamma_i -
(\alpha_i\Delta/\pi) (v_1 + v_2) \Bigr)
\nonumber \\
& = & \left[\frac{\pi}{\alpha_i\Delta f(r)} \right]\ \exp
\left( -\frac{\pi\gamma_i}{\alpha_i\Delta[1-f^(r)^2]} \right) \sinh
\left( \frac{\pi\gamma_i f(r)}{\alpha_i\Delta[1-f(r)^2]} \right)
\ .
\label{eq:e4.6}
\end{eqnarray}
The limit $r\gg\lambda$ (points placed far apart) can be easily
extracted from Eq. (\ref{eq:e4.6}) and it agrees with Eq.
(\ref{eq:e4.2}). For the opposite limit, $r=0$, when the points
coincide, one gets
\begin{equation}
P_2(\gamma_i) = \left( \frac{\pi}{2\alpha_i\Delta}\right) \exp
\left( - \frac{\pi\gamma_i}{2\alpha_i\Delta} \right) \ ,
\label{eq:e4.7}
\end{equation}
which is the distribution for a pointlike lead with a coupling
constant twice as large, as we would expect.

The large fluctuation tail of the distribution tends to an exponential
function, just as Eq. (\ref{eq:e3.1}). The correlation between points,
however, renormalizes the coupling constant $\alpha_i$ to
$\alpha_i[1+f(r)]$. Notice also the linear behavior at
$\gamma_i\rightarrow 0$, {\it for any} $r\ne 0$, consistent with what
we anticipated. The slope of the distribution at small widths becomes
steeper as the distance between points decreases, but the qualitative
aspect is rather independent of $f(r)$ and one can usually approximate
$P_2(\gamma_i)$ by Eq. (\ref{eq:e4.2}).

We stress that the main effect of point correlation within the lead is
that the characteristic number of channels of a lead is not quite
accurately given by the cross section divided by the electron
wavelength. We can say that there is a crossover between single and
double-channel behavior driven by the distance (correlation) between
the points, i.e., by the parameter $f(r)$.

In Fig. 2 we show some distributions of level widths for two-site
leads with different site separations obtained from our numerical
simulation. One observes that the correlation between sites is small,
even for $r=1$ (in lattice units) and at the low energy portion of the
spectrum (where $\lambda$ is larger than at the middle of the
spectrum). This effect, peculiar to the tight-binding model adopted
here, can also be visualized in Fig. 3, where we have plotted the
density autocorrelator as a function of site separation.

Given Eq. (\ref{eq:e4.6}) we can proceed to evaluate $R_2(g_m)$.  In
analogy to the pointlike case, an analytical treatment is viable only
for the two temperature regimes $T\ll\alpha_{1,2}\Delta$ and
$\alpha_{1,2}\Delta\ll T\ll\Delta$; for $T$ comparable to
$\alpha_{1,2}\Delta$, one has to compute the distribution numerically
and the dependence on the temperature cannot be rescaled out of the
distribution. For simplicity, we will concentrate on the analytical
results. We first present the expression we obtained for the resonance
conductance distribution in the $T\ll\alpha_{1,2}\Delta$ regime:
\begin{eqnarray}
R_2(g_m) & = & \frac{\theta(1-g_m)}{4f(r)^2 \sqrt{1-g_m}} \left\{
\frac{[1+f(r)^2][1+(2-g_m)(a^2-1)]}{[1+g_m(a^2-1)]^2}
\right. \nonumber \\ &  & - \
\frac{[1-f(r)^2][1+(2-g_m)(a_+^2-1)]}{2[1+g_m(a_+^2-1)]^2}
 -  \left.
\frac{[1-f(r)^2][1+(2-g_m)(a_-^2-1)]}{2[1+g_m(a_-^2-1)]^2}
\right\} \ ,
\label{eq:e4.8}
\end{eqnarray}
where
\begin{equation}
a_{\pm}=(a\pm f(r)\sqrt{a^2-1})/\sqrt{1-f(r)^2} \ ,
\label{eq:e4.85}
\end{equation}
and $f(r)$ is defined in Eq. (\ref{eq:e4.5}). In Fig. 4 we show
$R_2(g_m)$ for different values of $f(r)$ and the asymmetry parameter
$a$. Notice that the most probable value of $g_m$ is maximum when the
leads are symmetric, as we have argued in the previous section.

Clearly, by setting $r=0$ in Eq. (\ref{eq:e4.8}) we recover Eq.
(\ref{eq:e3.2}). In the opposite limit, $r\gg\lambda$, we can expand
all terms to $O(f^2)$ and obtain
\begin{equation}
R_2(g_m) = \frac{3g_m \theta(1-g_m)\left\{[1+8a^2(a^2-1)]+2g_m(a^2-1)
(1-4a^2)+g_m^2(a^2-1)^2 \right\}}{4 \ \sqrt{1-g_m} \
[1+g_m(a^2-1)]^4} \ .
\label{eq:e4.9}
\end{equation}

The distribution of conductance peaks in the regime
$\alpha_{1,2}\Delta\ll T\ll\Delta$ can again be expressed as a
universal, temperature independent function if one uses the rescaling
shown in Eqs. (\ref{eq:e3.4}) and (\ref{eq:e3.5}). After some
straightforward manipulation we obtain
\begin{equation}
{\cal R}_2(x) = \frac{x^2}{4 f(r)^2}
\sum_{p=0}^3 (-1)^p e^{-a_px_p}[K_0(x_p) + a_pK_1(x_p)] \ ,
\label{eq:e4.10}
\end{equation}
where $x_0=x/[1+f(r)]$, $x_2=x/[1-f(r)]$, $x_{1,3}=x/\sqrt{1-f(r)^2}$,
$a_{0,2}=a$, $a_1=a_+$, and $a_3=a_-$. In Fig. 5 we have plotted
${\cal R}_2(x)$ for different values of $f(r)$ and $a$. When we set
$r=0$ in Eq. (\ref{eq:e4.10}), we recover Eq. (\ref{eq:e3.4}) with
coupling constants twice as large. On the other hand, in the limit
$r\gg\lambda$, one has, after an expansion to $O(f^2)$,
\begin{equation}
{\cal R}_2(x) = \frac{x^2 e^{-ax}}{2} \left[ x(a^2+1)K_0(x) +
(2ax+2a^2-1)K_1(x) \right] \ .
\label{eq:e4.11}
\end{equation}

In both temperature regimes, the most notable difference between
$R_2(g_m)$ and $R_1(g_m)$ is in the small $g_m$ behavior:
$R_2(g_m\rightarrow 0)$ is linear in $g_m$ {\it for any} $r\ne0$, in
contrast to $R_1(g_m)$, which tends to a constant in the same limit.
Correlations between points within the leads do not affect the linear
dependence.

\section{$N$-point leads}
\label{sec:n-site}

Some of the conclusions we have drawn for the two-point lead are
straightforward to extend to the $N$-point case. For instance, let us
call $Q_N(v_1,v_2,... ,v_N)$ the joint probability distribution of the
amplitudes $v_k=V|\psi(r_k)|^2$ at $N$ points. Then, since
$Q_N(0,0,...  ,0)$ is nonsingular for any arrangement where points do
not coincide (i.e., when no two points are completely correlated), we
{\it always} have $P_N(\gamma_i\rightarrow)\propto
\gamma_i^{N-1}$.

Unfortunately, the supersymmetry technique becomes impractical to use
in any derivation where $N>2$. We do not know of any other method
suitable for this task either. Therefore, in this section we will
simply assume that the point fluctuations within the leads are
completely independent, which is certainly correct for small enough
values of $\gamma_i$ or $g_m$. We will then derive expressions for
$P_N(\gamma_i)$ and $R_N(g_m)$ and look at the numerical simulations
to gain some insight about the general case, when correlations can be
present.

If the amplitudes $v_k$ fluctuate independently, we are allowed to
convolute $N$ distributions given by Eq. (\ref{eq:e3.1}) to obtain the
distribution of level widths for a $N$-point lead. In this way, we
find that
\begin{equation}
P_N(\gamma_i) = \frac{1}{\Gamma(N)} \left(\frac{\pi}{\alpha_i\Delta}
\right)^N \gamma_i^{N-1} \ \exp\left(-\frac{\pi\gamma_i}
{\alpha_i\Delta}
\right)  \ ,
\label{eq:e5.1}
\end{equation}
where $\gamma_i=(\alpha_i\Delta/\pi)\sum_{k=1}^N v_k$, and, according
to our definition, $\langle\gamma_i\rangle=N\alpha_i\Delta/\pi$. In
the limit $N\gg1$ one can check that the distribution becomes Gaussian
around $\gamma_i\approx\alpha_i\Delta N/\pi$, with a width
proportional to $1/\sqrt{N}$,
\begin{equation}
P_N(\gamma_i) \approx \frac{1}{\sqrt{2\pi N}}\left(\frac{\pi}
{\alpha_i\Delta} \right) \exp\left(-\frac{(\gamma_i-
\alpha\Delta N/\pi)^2}{2\alpha^2\Delta^2 N /\pi^2}\right) \ .
\label{eq:e5.2}
\end{equation}

The results of our simulations for $N=$4, 9 and 16 are shown in Fig. 6
for two different lead geometries and for extended eigenstates around
the bottom of the energy band. The deviation from Eq. (\ref{eq:e5.2})
becomes substantial for large $N$ when the lead sites are close
together. This indicates that the modest site-to-site correlation (see
Fig. 3) can add up to make the effective number of channels smaller
than the number of sites, particularly for small widths. As we move
the sites farther apart, correlations become negligible and the
distributions start to agree with Eq. (\ref{eq:e5.2}).

{}From Eq. (\ref{eq:e5.1}), together with Eqs. (\ref{eq:e1.6}),
(\ref{eq:e1.8}), and (\ref{eq:e1.9}), we can proceed to evaluate
$R_N(g_m)$. In the very low temperature regime,
$T\ll\alpha_{1,2}\Delta$, we obtain
\begin{eqnarray}
R_N(g_m) & = &
\frac{(g_m/4)^{N-1}\Gamma(2N)\ \theta(1-g_m)}{2\Gamma(N)
\sqrt{1-g_m}[1+g_m(a^2-1)]^{2N}}
\sum_{k=0}^{N}
\frac{(2N)!a^{2(N-k)}[(a^2-1)(1-g_m)]^{k}}{(2N-2k)!(2N+2k)!} ,
\label{eq:e5.3}
\end{eqnarray}
which contains all the essential features of $R_1$ and $R_2$. In the
other regime, where $\alpha_{1,2}\Delta\ll T\ll\Delta$, we obtain,
after an appropriate rescaling,
\begin{equation}
{\cal R}_N(x) = \left(\frac{x}{2}\right)^{2N-1}\frac{(2N)!e^{-ax}}
{\Gamma(N)^2}\left\{ \frac{K_0(x)}{(N!)^2} + \sum_{l=1}^{N}
\frac{[(\alpha_1/\alpha_2)^{l/2} + (\alpha_2/\alpha_1)^{l/2}]K_l(x)}
{(N-l)!(N+l)!} \right\} \ .
\label{eq:e5.4}
\end{equation}
One can easily check that Eq. (\ref{eq:e5.3}) becomes equivalent to
Eqs. (\ref{eq:e3.2}) and ({\ref{eq:e4.9}) for $N=1$ and $N=2$,
respectively. Similarly, Eq. (\ref{eq:e5.4}) is reduced to Eq.
(\ref{eq:e3.4}) for $N=1$, and to Eq. (\ref{eq:e4.11}) for $N=2$. We
remark that both Eqs. (\ref{eq:e5.3}) and (\ref{eq:e5.4}) yield the
same limit $R_N(g_m\rightarrow 0) \propto g_m^{N-1}$, i.e., the
distributions behave like a power law for small peak heights.

The large $N$ limit makes the distribution of conductance peaks
narrower. For instance, when $N\gg1$, Eq. (\ref{eq:e5.4}) tends to a
Gaussian law, centered around $x\sim O(N)$,
\begin{equation}
{\cal R}_N(x) \approx \frac{a}{\sqrt{4\pi N}}
\exp\left(-\frac{(x-2N/a)^2}{4N/a^2} \right) \ .
\label{eq:e5.5}
\end{equation}
Because $\langle g_m\rangle\sim O(N)$, but $\delta g_m\sim
O(\sqrt{N})$, fluctuations in the height of resonance peaks are
suppressed for very large leads. This should be contrasted to the
universal fluctuations of conductance as a whole, which obey $\langle
g_m\rangle/\delta g_m\sim O(1)$. Furthermore, $\langle
g_m\rangle\propto (1/\alpha_1+1/\alpha_2)^{-1}$, which is the
classical result (Ohm's law).

\section{Conclusions}
\label{sec:conclusion}

We have considered the statistics of the widths of electronic states
inside a quantum dot coupled to a bulk reservoir through contact
leads. It is well-known that in the case of pointlike contacts the
distribution of widths obeys a universal exponential law, which
corresponds to Gaussian fluctuations of the wavefunction. We have
shown that the statistics of the widths is strongly influenced by the
size of the contact area between the leads and the dot. For two-point
contacts we presented the exact expression for the distribution
function of widths and found that it closely resembles the result one
obtains by assuming no correlation between the points. The correlation
effects appear only for large fluctuations, whose contribution to any
observable quantity is exponentially small. Following this
observation, we suggested that the $N$-point lead can be considered as
a set of $N$ points fluctuating independently. We evaluated the
distribution of widths analytically and compared the results with
direct measurements from numerical simulations. A good qualitative
agreement was found.

The central region of the distribution tends to be Gaussian as $N$
increases. Therefore, the fluctuations in the level width are
suppressed for large leads and we can describe the dot in terms of the
average parameters. In the region of very large widths, the
distribution does not follow a Gaussian law, but rather decays
exponentially. For small widths, the distribution behaves as a power
law.

For the continuous system with extended leads we can expect a similar
behavior. The distribution of widths is Gaussian around the average,
which is proportional to the effective number of channels
$N_0=A/\lambda^{d-1}$, where $A$ is the cross section of the lead
contact. The tail of the distribution for very large widths is
exponential.  In the region of small widths the distribution behaves
as
\begin{equation}
P(\gamma_i) \propto \gamma_i^N \ .
\label{eq:6.1}
\end{equation}
The effective number of channels $N$ is equal to the number of
independent ``points'' in the lead, which increases for small
$\gamma_i$ as $N\propto N_0/\gamma_i$.

The relevance of these results to experiments appears in the
determination of the distribution of conductance peaks in the
resonance regime. The main consequence of the finite extension of the
leads is found in the power law behavior of the distribution at small
peak heights. As the lead becomes very large, the distribution tends
to a Gaussian law. The possibility of asymmetry between the leads is
taken into account in our work. Finally, we point out that Eqs.
(\ref{eq:e4.8}) and (\ref{eq:e4.10}) can be used to describe the
statistics of four-probe dots where the constrictions are made very
narrow (pointlike leads).

\begin{center}
{\large Acknowledgments}
\end{center}

We would like to thank A. Andreev, M. Faas, K. Efetov, C. Lewenkopf,
C. Marcus, and B. Simons for useful discussions. E. R. M. acknowledges
the financial support of Conselho Nacional de Desenvolvimento
Cient\a'{\i}fico e Tecnol\'ogico (CNPq, Brazil). This work was
supported through a NSF Grant No. DMR 92-04480.

\begin{figure}
\caption{Histogram of level widths for single-site
leads at two regions of the energy band: middle (circles) and bottom
(squares). The dashed line is the universal prediction (exponential
law). The level width is rescaled so that $\langle\gamma_i\rangle=1$.}
\label{fig:f1}
\end{figure}

\begin{figure}
\caption{Histograms of level widths for two-site leads at two regions
of the energy band: middle (squares) and bottom (circles). The level
width is rescaled so that $\langle\gamma_i\rangle=2$. $r$ is the
distance in lattice units between sites within the lead. The solid and
dashed curves are the predictions for $f(r)$=0.35 and $f(r)=0$,
respectively.}
\label{fig:f2}
\end{figure}

\begin{figure}
\caption{Density-density correlator as a function of site separation
for two regions of the energy band: middle (circles) and bottom
(squares). The density is defined as $\rho(r)=|\psi_\nu(r)|^2$ and the
average is performed over $\nu$ (eigenstate) and grid location.}
\label{fig:f3}
\end{figure}

\begin{figure}
\caption{Distribution of resonance conductance peaks in a dot with
two-point leads at very low temperatures ($T\ll\alpha_{1,2}\Delta$).
The main plot shows the curves for the symmetric case (a=1) and
different separation of lead sites. Notice the liner behavior as
$g_m\rightarrow 0$ for any f$\ne$1. The insert shows the dependence
with the asymmetry parameter for a fixed distance between sites
(f=0.5).}
\label{fig:f4}
\end{figure}

\begin{figure}
\caption{ Distribution of resonance conductance peaks in a dot with
two-point leads at intermediate temperatures ($\alpha_{1,2}\Delta\ll
T\ll\Delta$). The conductance is expressed in the rescaled form
$x=4Tg_m/\protect{\sqrt{\alpha_1\alpha_2}}\Delta$. The main plot shows
the curves at a=1 (symmetric case) and variable distance between the
lead sites. The insert shows the dependence on the asymmetry parameter
for a fixed f=0.5.}
\label{fig:f5}
\end{figure}

\begin{figure}
\caption{Histogram of level widths for $N$-site leads and for
eigenstates at the bottom of the energy band. The level width is
rescaled so that $\langle\gamma_i\rangle=N$. The dashed line is the
prediction for independent site fluctuations of the wavefunction.  The
squares correspond to leads where sites were placed together ($r=1$)
and the circles correspond to leads with $r=3$. As an example, in the
insert we show how a $N$=4 lead with $r=2$ was implemented.}
\label{fig:f6}
\end{figure}

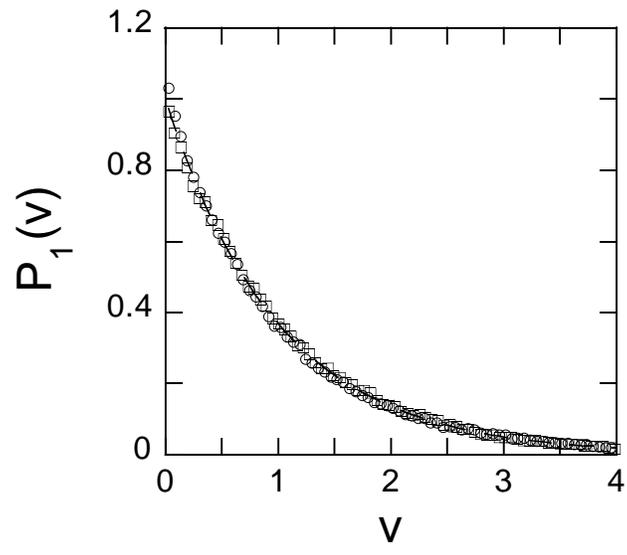
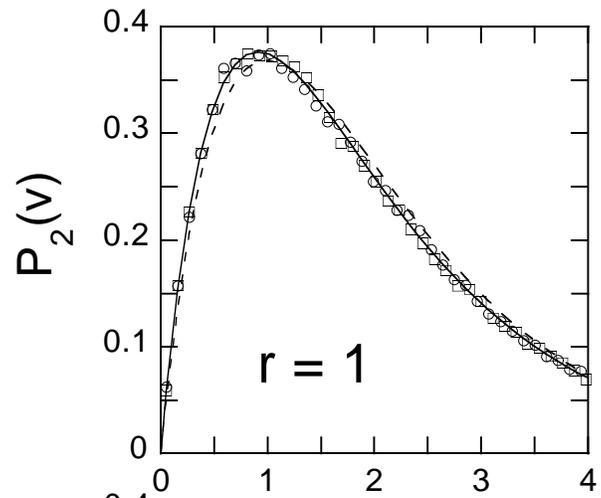
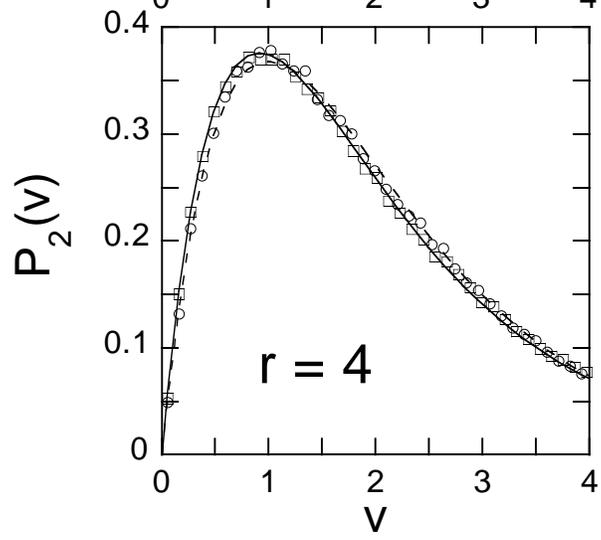
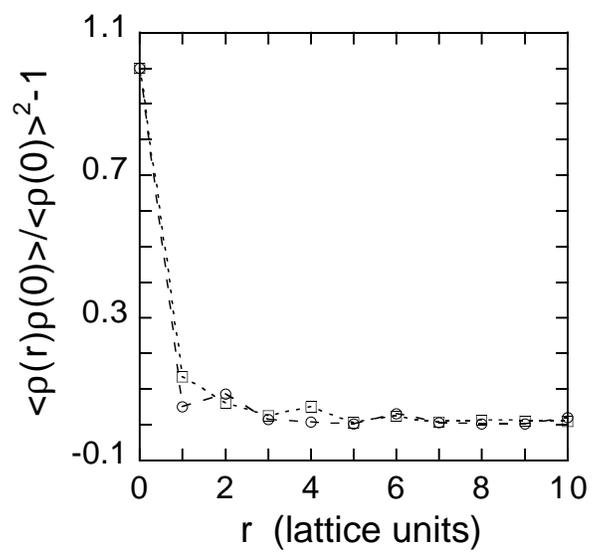
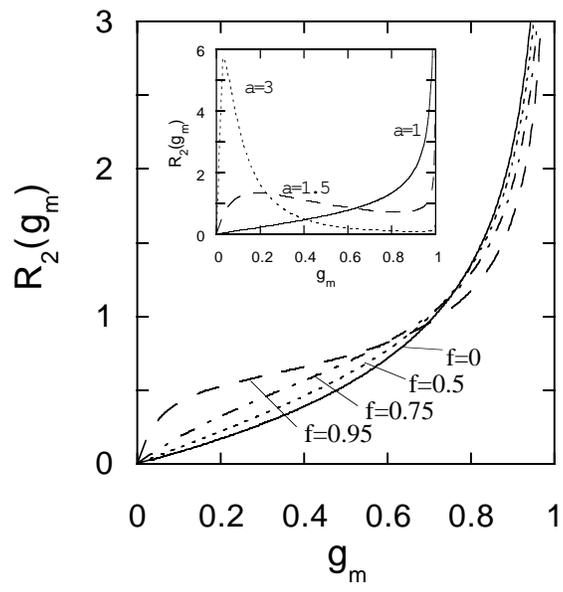
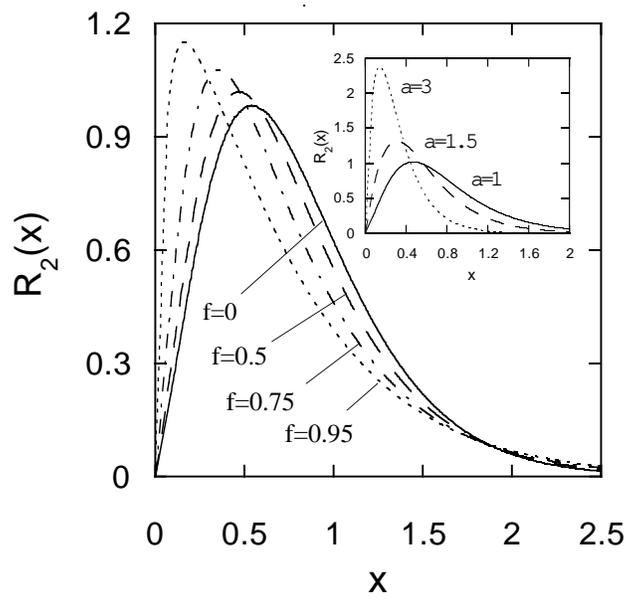
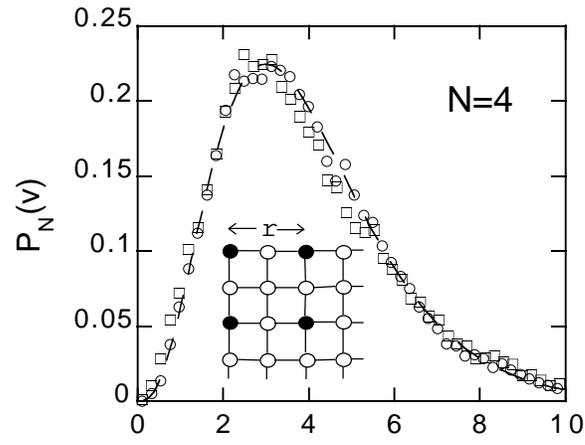
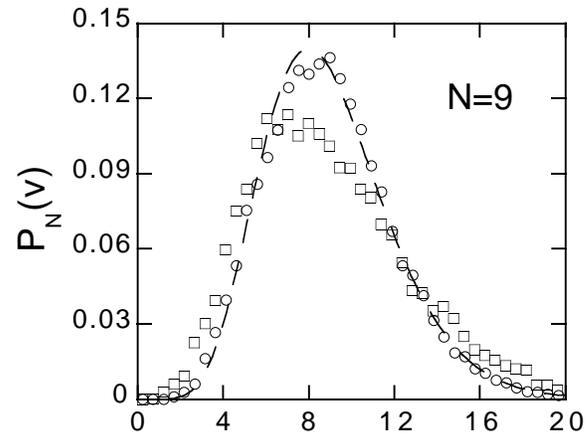
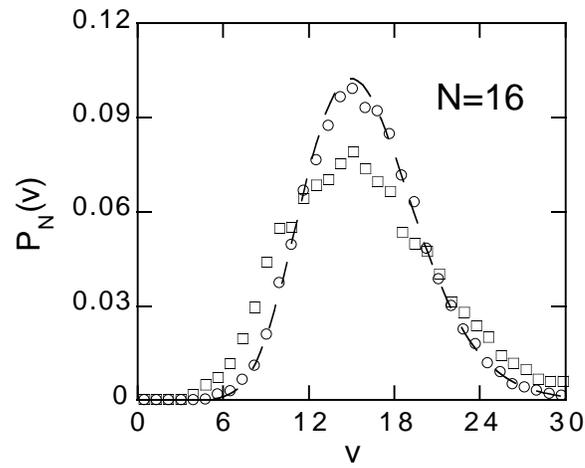
\end{document}